\documentstyle[11pt]{article}
\input epsf

\newcommand{\al}{{\alpha}}
\newcommand{\beq}{\begin{equation}}
\newcommand{\eeq}{\end{equation}}
\newcommand{\beqa}{\begin{eqnarray}}
\newcommand{\eeqa}{\end{eqnarray}}
\newcommand{\bi}{\begin{itemize}}
\newcommand{\ei}{\end{itemize}}

\newcommand{\vk}{{\vec k}}

\newcommand{\bra}[1]{{\left\langle #1\right|}}
\newcommand{\ket}[1]{{\left| #1 \right\rangle}}

 \title{Systematic Two-band Model Calculations of the GMR Effect 
with  Metallic and Nonmetallic Spacers and with Impurities\footnote{Based on
the {\it Diploma Thesis} of O.~Gebele, Regensburg 1998}}

\author{ O.~Gebele$^1\footnote{present adress: Gebele@tu-harburg.de}$,
 M. B\"ohm$^1$\footnote{present adress: mBoehm@wila-verlag.de},
 U. Krey$^1${\thanks{Corresponding author, FAX
(xx49) 941 943-4544, e-mail uwe.krey@physik.uni-regensburg.de}} \,\,and
S.~Krompiewski$^2$
 \\ $^1$
  Institut f\"ur Physik II der Universit\"at,
 93040 Regensburg, Germany
 \\ $^2$ Institute of Molecular Physics, P.A.N., 
 60-179 Pozna\'n, Poland                  \\
  }

\date{received: 12/07/1999; revised 3/12/1999; accepted 24/2/2000}

\input epsf
\begin{document}

\large
\maketitle
\begin{abstract}

By  a {{semi-empirical}} Green's function method we calculate conductances
and the corresponding Giant Magneto-Resistance effects (GMR) of two metallic
ferromagnetic films separated by different spacers, metallic and
non-metallic ones, in a simplified model on a $sc$ lattice, in CPP and CIP
geometries (i.e.\ current perpendicular or parallel to the planes), without
impurities, or with interface- or bulk impurities. The electronic structure
of the systems is approximated by two hybridized orbitals per atom, to mimic
s-bands and d-bands and their hybridization.

We show that such  calculations usually  give rough estimates only, but of
the correct order of magnitude; in particular, the predictions on the
impurity effects depend strongly on the model parameters. One of our main
results is the prediction of huge CPP-GMR effects for {\it non-metallic}
spacers in the ballistic limit.

\end{abstract}
\vspace*{0.5cm}

\section{Introduction}

 The presence of giant magneto-resistance effects in trilayer systems
consisting of two ferromagnetic metallic layers -- source and drain --
separated by a non-magnetic or anti-ferromagnetic crystalline metallic
spacer is meanwhile well-known \cite{Fert,Gruenberg}.
Recently also non-crystalline and non-metallic spacers
have been considered
 \cite{La1,La2,Buergler}, i.e.\ there is renewed  interest
in spin-polarized tunneling through semi-conducting or
non-conducting spacers \cite{Ted,Jul,Ste},
 based on  the perspective of new applications  in
magneto-electronic technology, e.g.\ for 
magnetic-field sensors, spin-valve transistors \cite{Mon} or spin-polarized
field-effect transistors \cite{Da1,Pri}.

 In \cite{Slo,Mat2} spin-valve properties of 
ferromagnet/insulator/ferromagnet junctions were studied, and 
\cite{But} presents a first $ab$ $initio$ 
 calculation of the {\it  one-particle} electronic and magnetic properties of FM/NM/FM
tunneling structures, with
FM=Fe, and NM=Ge and GaAs.

However concerning the GMR effect, which is based on {\it two-particle}
  properties, equally accurate calculations for FM/NM/FM systems,
particularly in the CPP geometry, apparently could only be performed without
impurities, \cite{Butler2}. For systems with {\it metallic}
 spacers one should mention at
the same time the calculations of Tsymbal and Pettifor, \cite{Tsymbal}, or
Mathon, \cite{Mathon97}, where also fully realistic calculations have been performed,
but again with phenomenological assumptions for the impurity scattering,
\cite{Tsymbal}, or with no impurities at all, \cite{Mathon97}. Furthermore,
in a two-band tight-binding approximation, two of the present authors,
\cite{KroKrey98}, have already treated systems, where one or two of the
ferromagnet-spacer interfaces were decorated with ultrathin non-metallic layers,
but again with translational invariance within the planes.
 In this approximation, the two bands mentioned above describe the s- and
d-electrons and their hybridization, or the conduction and valence band in
case of a non-metallic spacer.

 In the ferromagnetic metals, the parameters
of the d-band are of course spin-dependent.

Since in \cite{KroKrey98} the method for the calculation of the
resistivities turned out to be extremely flexible and accurate, we extend it
in the  present
paper to a systematic survey of the magnetoresistivity of
trilayers of two ferromagnetic sandwiches separated by different
non-magnetic spacers, metallic and non-metallic ones, with impurities of
various kind, and
without impurities, in CIP and and CPP geometries, but within a 
{{simplified  semi-empirical hybridized two-band
model}}.

We stress that our approach is rigorous in principle,   numerically very
 accurate, and avoids the Coherent Potential Approximation (CPA),
 which is often applied to disordered systems, but difficult to control.

As already mentioned, there exists already a large body of theoretical
calculations on the GMR of magnetic multilayer systems.
For systems with metallic spacers and without impurities  our simplified two-band model
calculations, to be presented
below, can of course not { compete}  -- and this is  also not
at all intended -- with the much more ambitious and cumbersome calculations
mentioned above   e.g.\ those of \cite{Mathon97}, who performed fully
realistic calculations of the CPP- and CIP-GMR for Co/Cu/Co/Cu and
Fe/Cr/Fe/Cr multilayers with ideal leads made of Cu rsp.\ Cr, in the
ballistic, Ohmic, and Anderson-localized regimes corresponding to the cases of
 no-disorder rsp.\ certain cases of one-dimensional disorder, i.e.\ with
respect to the stacking of the planes, but without impurities. There exist
more such ambitious calculations, e.g.\ the seminal paper of Schep {\it et
al.}, \cite{Schep}, and other publications, which also in principle would
have earned citing but which are omitted on lack of space. In this respect
the purpose of our paper is, instead, to study in some detail the question,
{\it to which extent} simplified models as the present one are able (or not)
to reproduce those results.

 On the other hand, with impurities, on which we concentrate,
similar extensive calculations 
{{of corresponding rigour}
do not yet exist, to our knowledge. In fact,
concerning the influence of impurities, the 
{{\it methods}}
 of our calculation have
already been presented in the paper of Asano {\it et al},
\cite{Asano}. However, rigorous calculations with impurities are so
demanding that those authors, although they  wrote already down the formalism for a
{{\it two-band model}}, published numerical results only for
{{\it one-band}} cases.

Here we present results obtained with the two-band model within the
formalism of \cite{Asano}. For a {\it nonmetallic} spacer it is actually necessary
to consider at least two orbitals per atom to simulate the valence and
conduction bands, respectively, but also for ferromagnetic or non-magnetic
3d-{\it metals} it is a natural requirement to include a second band with the
intention to mimic as far as possible the d-electrons in addition to the
s-states. Namely, the d-electrons are not only responsible for the magnetic
properties, but also for a large part of the resistivity, due to
{{\it  Mott's
$s$-$d$-scattering mechanism}}. Therefore it is also important to include not
only two bands, but also realistic values for the $s$-$d$ {\it
hybridization}.

This should give enough motivation to our two-band calculation; but  we
stress again that these calculations should be considered as model
calculations and are not intended to replace fully realistic calculations,
as far as they are possible already now, e.g.\ \cite{Mathon97}, or in future.
In fact, it turns out that the present two-band model calculations will be
able to give estimates of the GMR, which yield the correct order of
magnitude, but not more. Further, concerning the impurity effects, it turns
out that certain contradictory results (see below) concerning essential
trends depend strongly on the model-parameters, so that again, now with
impurities, completely realistic calculations cannot be avoided: Unfortunately, such
 calculations, where the impurities are treated with the same kind
of rigour as the electronic structure of the pure system itself, are
apparently not yet possible,
{{\cite{Mertig1}--\cite{Mertig4}}}.

 On the other hand, one of our most remarkable predictions, which should not
be overlooked and can be stated already from the present model
calculations, is that a large enhancement of the CPP-GMR may happen, if
{\it nonmetallic spacers} of increasing thickness are used. Here it is
{{of course}}
required that the thickness of the nonmetallic spacer remains smaller than
the dephasing length for inelastic scattering; i.e.\ one should remain in
the ballistic limit.

In the following sections we present the basic theory and then the results
of our\, survey, which is followed at the end
 by a section presenting our conclusions.
\section{Basic definitions}
\subsection{Systems}
Our systems are defined in Fig.~1 and Fig.~2 for the CPP and CIP geometries,
respectively.

We start with metallic 'ideal leads' on the left-hand side and
right-hand-side, respectively,  followed  -- in the CPP geometry -- by three monolayers
of ferromagnetic metal $F_1$ rsp.\ $F_2$, with $n_s$ spacer monolayers
inbetween.
In the CPP geometry all layers have a quadratic cross-section of $M\times M$
square-lattice unit cells.

For the CIP geometry, the natural definition of our systems is similar and
explained by Fig.\ 2.

The electronic structure of our systems is described by the following
tight-binding Hamiltonian\ :
\begin{eqnarray*}
 {\cal H}& = &\sum_{l,i}\,E_{l.i}^s\,\sum_{\sigma}
            \hat c^+_{l,i,\sigma}\hat c_{l,i,\sigma}
          +\sum_{l,i}\sum_{l',i'}t^s_{l,i,l',i'}
           \sum_\sigma\hat c^+_{l,i,\sigma}\hat c_{l',i',\sigma} \\
         & + &\sum_{l,i}\sum_{\sigma}\,E^d_{l,i,\sigma}\,
            \hat d^+_{l,i,\sigma}\hat d_{l,i,\sigma}
          +\sum_{l,i}\sum_{l',i'}t^d_{l,i,l',i'}
           \sum_\sigma\hat d^+_{l,i,\sigma}\hat d_{l',i',\sigma} \\
         & + &\sum_{li}\,V^{s,d}_{l,i}\,\sum_{\sigma}
            (\hat d^+_{l,i,\sigma}\hat c_{l,i,\sigma}+\hat c^+_{l,i,\sigma}
\hat d_{l,i,\sigma})\,\,.
\end{eqnarray*}

Here $\hat c^+_{l,i,\sigma}$ and $\hat c_{l,i,\sigma}$  denote creation and
destruction operators of an s-electron occupying the site $l$ in plane $i$
with spin $\sigma$. The corresponding operators for d-electrons are
$\hat d^+_{l,i,\sigma}$ and $\hat d_{l,i,\sigma}$. The
band-energy parameters are $E_{l,i}^s$ for the s-states, which do not depend on
the spin $\sigma$, and $E^d_{l,i,\sigma}$,
which are spin-dependent, for the d-states. The hopping-matrix elements for s-
and d-states  are
$t^s_{l,i,l',i'}$ and $t^d_{l,i,l',i'}$, and the local s-d-hybridization is
given by $ V^{s,d}_{l,i}$.

  Of course, taking more bands into account,
 i.e.\ {{all}} five d-bands, {{ the 4s-band and the
three 4p-}}{{bands, and adding the spin dependence}}, 
we could have tried  to fit the band
structure of realistic systems as far as possible, see e.g. \cite{Sanvito};
but this is 
{{\it not}} our purpose, since we try to remain  semi-quantitative and study
only the essential trends.
 Particularly, in view of the still much simplified band structure, we also
use typical simplifications in the description of the impurities, see below.
But in principle, within our formalism, more rigour in both respects
would be possible, but only at the cost of excessive computing.

For the following, $\cal H$ is written in matrix representation with respect
 to an orthonormal basis by replacing
the creation and annihilation operators by the corresponding
ket- and bra-vectors, respectively, such that ${\cal H}\to \hat
H$, e.g.\ by $\hat d^+_{l,i,\sigma}\hat c_{
l,i\sigma}\to \ket{d,l,i,\sigma}\bra{s,l,i,\sigma}$. 
\subsection{Resolvent operators and Kubo formula}
For the calculation of the conductances we use the formalism of
Fisher and Lee, \cite{Fis}. Therefore, we need the following
matrix elements of the advanced and retarded resolvent operators $\hat
 G^{\pm} :=\{ E \pm {\rm i}\,0^+ -{\hat H}\}^{-1}$~:
\begin{equation}
G^\pm(i,i')_{\al,l,\sigma\,;\,\al',l',\sigma'}=
\bra{\al,i,l,\sigma}\{E\pm {\rm i}\,0^+ -{\hat H} \}^{-1}
\ket{\al',i',l',\sigma'}\,\,.
\end{equation}
Here the index $\al$ counts the orbitals, e.g.~$\al = s$ or $d$.
As usual, we also call these matrices 'Green's operators', and ${\rm
i}\,0^+$
is a positive  imaginary infinitesimal, 
such that the matrix-inverse in this
equation always  exists in our approach.

 For auxiliary purposes we need the 'left-sided' and
 'right-sided' Green's operators $G^L$ and $G^R$, e.g.\
  \begin{equation}
G^L(i_0)_{\al,l,\sigma\,;\,\al',l',\sigma'}\, :=\,
\bra{\al,i_0,l,\sigma}\{E+{\rm i}\,0^+-\hat H_L^{i_0}\}^{-1}
\ket{\al',i_0,l',\sigma'}
\,\,.\end{equation}
Here $\hat H_L^{i_0}$ denotes the Hamiltonian for a system, in which all
planes $i > i_0$ are deleted; $G^R(i_0)$ is defined in a similar way with
$E+{\rm i}\,0^+-\hat H_R^{i_0}$, where $ H_R^{i_0}$
 is the Hamiltonian of a system with all 
planes $i < i_0$  deleted.
 $G^L$ and $G^R$ are obtained by {\it recursion}, e.g.\ $G^L$ by recursion
 from the left, i.e.\
\begin{equation}
G^L(i_0)=[\hat g(i_0)^{-1}-\hat TG^L(i_0-1)\hat T]^{-1}\,\,.
\end{equation} 
$\hat g(i_0)$ is  the resolvent of the isolated plane $i_0$; $\hat T$ is
the matrix describing the 'hopping' from plane to plane.

 From this one gets finally the two desired resolvent operators 
\begin{equation}
G(i,i)=[\hat g(i)^{-1}-\hat TG^L(i-1)\hat T-\hat TG^R(i+1)
\hat T]^{-1}
\end{equation} 
and
\begin{equation}
G(i,i+1)=G(i,i)\hat TG^R(i+1)\,,
\end{equation}
which are needed for the conductance, see below.
Here the plane $i$ ($=i_0$) is arbitrary, since under stationary
 conditions through any of the
planes the same amount of current is flowing.

Finally, with the definition
\begin{equation}
\tilde G(i,i')\,:=\frac{1}{2{\rm i}}[G^-(i,i')-G^+(i,i')] 
\end{equation}

one gets the conductance from the
{ Kubo-formula}, \cite{Fis}:

\begin{equation}\label{eqKubo}
\Gamma_\sigma=\frac{4e^2}{h}{\rm Tr}_{l\,,\,\alpha }
\,[\tilde G(i,i)\hat T\tilde G(i+1,i+1)
{\hat T}
-{\hat T}\tilde G(i,i-1)\hat T\tilde G(i,i-1)]\,,
\end{equation} 
where the trace ${\rm Tr}_{l\,,\,\alpha }$
is performed with respect to the orbitals $\al =s,d$,
and with respect to the sites $l$ of the atoms of a given plane, whereas
the spin-index $\sigma$  ($=\sigma'$) is kept fixed. Here $e$ is the
electronic charge, and $h$ is Planck's constant.

 The GMR effect is  defined by the equation
\beq {\rm GMR}\,=\,
\frac{\Gamma^{++}_{\sigma=\uparrow}+\Gamma^{++}_{\sigma=\downarrow}}
{\Gamma^{+-}_{\sigma=\uparrow}+\Gamma^{+-}_{\sigma=\downarrow}} -1
\,=\,\frac{\Gamma^{++}}
{\Gamma^{+-}} -1\,\,.
\eeq
Here the superscripts ($++$) resp.\ ($+-$) denote 
{{\it 
mutually parallel}} rsp.\ 
{{\it 
antiparallel}} magnetizations of the two ferromagnetic sandwiches. Since
 we neglect spin-flip scattering altogether, the conductances are of course
additively composed out of the separate contributions from majority and minority
spins, respectively, as denoted in the formula.

We have performed real-space (x-space) and k-space calculations. In the
x-space calculations each layer consists of $M\times M=$10$\times$10 atoms
with free boundary conditions, {{\cite{REMrealspace}}}.
 Here the matrices are of size 200$\times$200, again for given
$\sigma$ ($=\sigma'$). Impurities can be added at will, which is important
below.

In our $k$-space calculations, which are only performed in the CPP case
and without impurities, the planes are infinitely extended.
 Here, in the trace of equation (\ref{eqKubo}), in the $k$-space
 representation, the sum over $l$ is replaced by the corresponding
  sum over $\vec k$-vectors in the two-dimensional Brillouin zone,
since the two-dimensional vector $\vec {k}_{||}$ is now 'a good quantum number', i.e.\
there is translational invariance in the planes.   This summation, or
the corresponding integral, is approximated extremely
accurately by a Cunningham formula, \cite{Cun}, with $\sim 10^6$ 
$\vec{k}_{||}$-points, for which the matrix elements of the Green's operators
can be calculated separately without difficulty and extremely accurately. (All
numerical evaluations have been performed with  Cunningham's accuracy parameter
$m=10$ or $11$).
\subsection{Model parameters}
For the models considered, we have chosen four parameter sets: Two parameter
sets, (i) and (ii), correspond to a {\it metallic} spacer and mimic the
main features of trilayers composed of Co as ferromagnetic metals and Cu as
nonmagnetic metallic spacer material (see below), whereas the third and
fourth parameter sets, (iii) and (iv), correspond again to Co ferromagnets, but
with a {\it non-metallic} spacer.

Concerning the  sets (i) and (ii), we note that, as already mentioned, they are
constructed to reflect the two main features of Co and Cu as far as
possible in our simplified two-band model: The two features are
\bi \item the similarity of the s-bands of Co and Cu, and
\item the similarity of the d-bands of Cu and the 'majority-spin' d-bands of
Co.
\ei
This  qualitative 
similarity is obvious from  Fig.~3, which presents
results obtained by Mathon {\it et al.}, \cite{Mat3}.
 The parameter sets (i) and (ii) are in fact derived from rough
 fits to {\it the overall band-structure of Co} --
parameter set (i), see Fig.\ 4a --  respectively
 to {\it the overall structure of the DOS for Co} -- parameter set (ii), see
Fig.\ 4b.
But these  fits are rather crude anyway, and so the
derivation  is not emphasized. The values are given in
Table 1 and Table 2, respectively, \cite{REMgroupvelocity}.

 In Tables 3 and 4, the parameter sets (iii) and (iv)  are
given, which correspond
to {\it nonmetallic} spacers ('Isolator
 1 and 2', respectively). The parameters of the ferromagnet correspond to
case (i), with $E_F=-2.8$.

\noindent 
The notions 'indirect energy gap'  and 'direct
energy gap', which are used in the table captions of Table 3 and 4,
respectively, refer to Fig.~5:

\noindent For parameter set (iii) there is an '{\it indirect energy gap}', since the
 valence-band maximum is at $k_{[111]}=(\pi/a)\sqrt{3}$ with an energy -2.81,
which is slightly below the Fermi energy $E_F= -2.80$, whereas the minimum
of the conduction-band energy is at $k_{[111]}=0$ with E=-2.79. These
numbers refer to the first line of the three parameter sets of Table 3.

 On the other hand, with parameter set (iv), also with $E_F=-2.8$, there is
 a {\it direct} energy gap at $k_{[111]}= 0.814$,
 i.e.\ at the {\it arrow} in the r.h.s.\ of Fig.\ 5. In both figures, and
in the results presented below, the uppermost line, i.e.~with the smallest
gap, of Table 3 and Table 4 has been used.

\subsection{Impurity models}
We consider a) bulk impurities and b) interface impurities.

In case a), we have considered {\it bulk impurities} only in the spacer,
{{and exclusively nonmagnetic ones, which}} 
 are modelled by adding to the parameter $E_d$ of the single-site
d-band energy a {\it Gaussian spatial noise} proportional to a 'disorder
strength'  $\sigma_r$, namely
\beq \label{eqBulk_impurities} E_d\to E_d + \sigma_r\cdot n_l^{\rm Gauss}\,.\eeq
Here the $n_l^{\rm Gauss}$, where $l$ enumerates the different spacer atoms,
 are independent random numbers  distributed according to a 'Gaussian' with
average 0 and variance 1; 
the parameters of the s-bands remain unchanged. 

These assumptions for our bulk impurities are of course rather schematic,
i.e.\ these are the same simplified assumptions e.g.\ of Asano {\it et al.},
\cite{Asano}. But principally one might have been more ambitious and might
have considered e.g.\ specific magnetic impurities. Such calculations are of
course always possible within our formalism,
{{\cite{REMgeom,Garcia}}},
 and have partially been performed in
course of the diploma work of the first one of the present authors;
 however, as a first step, and since the results with the magnetic
impurities were not particularly exciting and have not been documented,
 also because of lack of space, we keep at present to our simplified
description of the impurities. This seems also justified in view of the
simplifications made for the band structure and for the geometry of our
systems.

Concerning case b), the {\it interface impurities} are defined as follows~:
For every ferromagnet-spacer interface, in each of the two boundary
layers of the interface there are $n$ atoms (out of $M\times M$) selected at
random and replaced by atoms of the other kind.

For impure systems we have of course {\it averaged}
 over a large number of different samples, see the figures below; this leads
to the 'averages with error bars' in the results presented in the following
section.

\section{Results}
\subsection{CPP-GMR}
 \subsubsection{CPP-GMR; no impurities}
 At first we describe the results obtained without impurities obtained
 in the CPP case ('current perpendicular to the planes'). In Fig.\ 6
and Fig.\ 7
the results of k-space calculations with model parameters
 (i) and (ii) are presented,
 for variable thickness $n_s$ of the 'metallic spacer'. Interestingly, 
in both cases there is a slight but significant spatial oscillation
of the GMR as a function of $n_s$;  in case (i) the GMR oscillates between 
$\sim 70$ and $\sim 80$ \% with a quasiperiod of $\lambda \approx 5$
 monolayers (ML); in case (ii)  the oscillation of the GMR is between $\sim 47$
and $\sim 51$ \% with $\lambda \approx 3$. Here one can explicitly see that
 the conductance is {\it ballistic} from the fact that on average the
conductances hardly depend on $n_s$ in the limit $n_s\to\infty$; in
contrast, in case of {\it Ohmic} behaviour, one would expect that the
conductances  converge to 0, namely $ \propto n_s^{-1}$, as soon as
$n_s$ becomes larger than the relevant ''scattering length'' for diffusive
elastic or inelastic scattering, \cite{Mathon97}.

The reason for the spatially oscillating behaviour of the GMR are tiny 
oscillations of the conductances $\Gamma^{++}_\uparrow$ and
$\Gamma^{++}_\downarrow$, which are visible in the plots on the r.h.s.\
 of the preceding figures. These oscillations are caused by our treatment of
the ideal leads: In fact, for the ideal leads we have always used the same
parameters as within the 'Cu' spacer, but usually -- with exceptions mentioned
below -- for the pure system the s-d-hybridization has been neglected in the
 ideal leads (but not in the spacer). The reason for this neglection  is
mainly technical, since in case of neglected hybridization we can
directly express the matrix elements of the Green's operators $G^L(\vk, i)$
 and $G^R(\vk,i)$ at the boundary layers from the ideal lead to the 
ferromagnetic by simple {\it analytical} expressions, whereas the
 numerical calculation
 is much harder if the s-d-hybrization is taken into account already
in the ideal leads, by which the above-mentioned tiny oscillations vanish.
 What is, however, more important:
 In this way we have made explicit that the properties
of the ideal lead really matter, which is not astonishing with
 'ballistic electrons', where the 'contact resistance' from the ideal
leads to the system under study plays an important role.

\noindent{\it
As a first consequence one can learn from this study that it is a mistake
to speak of 'the ideal lead' as a uniquely defined entity:
In ballistic experiments there are {\rm
different} ideal leads, and it is necessary to characterize them, too,
as already stressed in a recent book on electronic transport in mesoscopic
systems}, \cite{Da2}.

It is also interesting at this place to contrast the different, i.e.\
parameter-dependent, results of our Fig.~6 and Fig.~7 with the corresponding 
results found by the much more ambitious {\it ab initio} calculations of
\cite{Mathon97} and \cite{Schep}. These authors find values of the CPP-GMR
for Co/Cu-multilayers amounting to $\sim 150$ \%.
 Moreover, for the conductances $\Gamma/M^2[e^2/h]$, instead of our
values 0.8; 0.3; and 0.2 (Fig.\ 6) rsp.\ 0.45; 0.2; and 0.15 (Fig.\ 7) for
the cases (++,$\uparrow$), (++,$\downarrow$), and (+-), respectively, they
find values of
$\Gamma$ corresponding to 0.43$\cdot$10$^{15}$, 0.25$\cdot$10$^{15}$, and
0.165$\cdot$10$^{15}$
$\Omega^{-1}$ $m^{-2}$, i.e.\ $\Gamma/M^2[e^2/h] = 0.67$, 0.39,
and 0.26.  

{As a conclusion, our values are correct concerning the order of the three
conductances, and their order of magnitude, but not more. This is exactly what
can be expected.}
\subsubsection{CPP-GMR; bulk impurities}
We would  like to stress at this place
 that with impurities, all calculations have been
performed in the {\it real} lattice space, with cross sections of
$M\times M=$10$\times$10 sites, {\it free} boundary conditions, in the
perpendicular directions,
{{instead
of {\it periodic} ones}}, and with complete
s-d-hybridization everywhere, i.e.\ now always {\it including} the ideal leads.

In  Fig.~8 and Fig.~9, for $n_s=3$ and the parameters (i) and (ii),
respectively, our results for the GMR and the conductances are presented over the
strength of the disorder, i.e.~over the standard deviation
 $\sigma_r$ in equation (\ref{eqBulk_impurities}). For the parameter set
 (i), Fig.\ 8, one obtains first a rapid {\it decrease} of the GMR from GMR $\approx
 1.15$   to ${\rm GMR} \approx 0.45$, when  $\sigma_r$
 increases from 0 to 1, followed
  by  a plateau behaviour with further increase of $\sigma_r$. Concerning
  the  behaviour of $\Gamma_\sigma^{++}$,  $\Gamma_\sigma^{--}$, and 
  $\Gamma_\sigma^{+-}$, one can see from the r.h.s.~of Fig.~8 that the
  GMR reflects essentially the behaviour of   $\Gamma_\uparrow^{++}$.

For the parameter set (ii), Fig.\ 9, there is also at first a rapid decrease
 from ${\rm GMR} \approx 0.37$ at $\sigma_r =0$ down to ${\rm GMR}\approx 0.15$
at $\sigma_r \approx 1$, but with a further increase of $\sigma_r$ there is
now a significant {\it re-increase} of ${\rm GMR}$
 back to $\sigma_r \approx 0.22 $ for $\sigma_r
\approx 6$. In fact, in this case  the behaviour of the conductances,
which are plotted on the r.h.s.\ of Fig.\ 9, is significantly
different and more subtle than that of Fig.\ 8.  In particular, the results
of Fig.\ 9 differ from what in {\cite{Asano}} the authors have obtained from a
simple one-band model.

 {\it So from these two figures, Fig.\ 8 and Fig.\ 9, concerning the
dependence on  the strength of disorder $\sigma_r$, one may only conclude that at first a
rapid decrease of ${\rm GMR}(\sigma_r)$ from its value at $\sigma_r=0$ to
 ${\rm GMR}(1)\,\approx 0.4\cdot {\rm
GMR}(0)$
 would be typical; this qualitative behaviour
seems to be essentially
parameter-independent; but for $\sigma_r >1$ a quantitative
calculation is hardly avoidable}, \cite{REMgmr,Moser}.

The reason for the common '{\it decreasing behaviour}'
 for $\sigma_r\widetilde < 1$ is the
following: In this case, an increase of $\sigma_r$ leads to an effective
reduction of the number of  open channels for coherent motion of
an electron from one plane to the next. This reduction concerns mainly
$\Gamma^{++}_\uparrow$ and therefore leads to a {\it decrease} of the GMR, since
the majority-spin d-electrons, {\it without} impurities, would not feel any
potential-variation at all. 

 In this connection, a very recent paper of Bruno {\it 
et
al.}, \cite{Bruno98}, should be mentioned: There it is shown that
 the conductance of
systems with impurities can be decomposed into two parts, a '{\it
diffusive}' part, arising from scattering processes, where the in-plane $\vec
k_{||}$-vector is not conserved, as is generically always the case for the present
systems with impurities and free-boundary conditions,
 and a '{\it ballistic}' part, to which only those processes
contribute, where the $\vec k_{||}$-vector is the same for
the incoming and outgoing waves. The diffusive part ( rsp.\ ballistic part)
always dominates the conductance, as long as the spacer length is much
shorter (rsp.\ much longer) than the {\it scattering length} $l_s$ of the
system. According to \cite{Bruno98}, $l_s$ should be of the order
of several 100 lattice spacings under similar conditions as in the present
work. So in our systems
with impurities, where $n_s$ is very small ($n_s=3$),
 always the {\it diffusive} part of the conductance
dominates.

\subsubsection{CPP-GMR; interface impurities}
In the following Fig.~10 and Fig.~11 the number $n$ of impurities
is varied between $n=0$  and $n=80$ for every interface layer
(remember that per interface, two layers are involved, with 100 sites per
layer; so we have $2\times n$ impurity sites out of 200 per interface, occupied by a
'wrong atom').

For the parameter set (i), Fig.\ 10, the ${\rm GMR}$ {\it decreases} with
increasing number of impurities, whereas for parameter set (ii),
Fig.~11, there is again the opposite behaviour: Here the ${\rm GMR}$
{\it increases} with increasing number of impurities.

\noindent This difference contains again an important statement in itself, which
should be particularly relevant to the experimentalist\  :

{\it Without a
precise calculation for a specific model one cannot predict whether an
{\rm increase} of the number of interface impurities will lead to an {\rm
increase} or a {\rm decrease} of the ${\rm GMR}$.}

\subsection{CIP-GMR}
\subsubsection{CIP-GMR; no impurities} The formalisms of the preceding
sections can also be applied on equal footings to the situation, where the
current is in the direction of the planes (CIP geometry, see Fig.~2). The
ideal leads have cross-sections of $\Delta z\times \Delta y=N\times M$
with
$N=6+n_s$, see below, while the CIP-conductor consists again of $\Delta z=N$
 monolayers of length $L=\Delta x=10$ and width $\Delta y=M=10$, namely two
ferromagnetic sandwiches of heigth $\Delta z=3$ and inbetween a spacer
sandwich of $\Delta z=n_s$. Again we consider at first the situation without
impurities; however in this geometry we have applied the x-space calculation
right from the beginning, i.e.~with cross sections of size $10\times 10$ and
with {{\it free}} boundaries, although with respect to the coordinate $y$
one could have still performed a Fourier transform. Moreover, as for the CPP
geometry without impurities, this time the s-d-hybridization has been
switched off for the lead wires.

 In Fig.~12 and Fig.~13, the ballistic CIP conductances are plotted as a
 function of $n_s$, for the parameter sets (i) and (ii), respectively.
In the first case, with increasing $n_s$ the conductances $\Gamma/M^2$
 rise, because the spacer cross section increases,
 the limiting value given by the ''shunting'' through the Cu spacer. The GMR shows 
irregular oscillations as a function of $n_s$; these oscillations change
with the boundary conditions, the geometry, and the material,
  and the results are more than one order of magnitude smaller than for the
CPP geometry, \cite{REMcip,Gijs,Theeuwen}.
Note that with parameter set (i), the GMR is even  negative on average over
$n_s$, which is not observed experimentally. By discrete Fourier
transformation of GMR  with respect to $n_s$, \cite{Gebele}, we would get
maxima in the Fourier spectrum at periods of 2.5, 4.6 and 8.33 monolayers
(ML) from parameter set (i), Fig.~12, and 2.27 and 5 ML from (ii), Fig.~13.

\subsubsection{CIP-GMR; bulk impurities} 
In case of bulk impurities, we have considered a system with $n_s=4$, i.e.\
$N=L=10$. The results are presented as a function of $\sigma_r$, the
 parameter characterizing the 'strength of the disorder' according to
 Eq.~($\ref{eqBulk_impurities}$). Here, in case of the parameter sets (i)
and (ii), we find again quite different behaviour in Fig.\ 14 and 15,
respectively: In case of Fig.\ 14, i.e.\ for parameter set (i), the GMR
{\it increases} at first from 0 to $0.015$, and then for $\sigma_r > 1$ it
roughly  remains constant with increasing $\sigma_r$, whereas in case of
the parameter set (ii), i.e.\ in Fig.\ 15, there is at first a drastic
{\it decrease} from GMR $\sim 0.035$ to GMR $\sim 0$, and then again a roughly
constant value of GMR $\sim 0$, {{\cite{REMcip}}}.
 In both cases the {\it conductances} decrease at first rather fast, and
then there is a slow re-increase; but this behaviour is slightly different
for the up- and down-spin channels in the ($++$)-configuration, and for the
($+-$)-configuration.

 {\it So again the behaviour of the GMR as a function of $\sigma_r$ is
 hardly predictable without an extensive and completely realistic
 calculation for a specific model.}
\subsubsection{CIP-GMR; interface impurities}
A somewhat different conclusion refers to the influence of {\it interface
impurities} for the CIP geometry.  Here in Fig.\ 16 and 17 we use the same
geometry as in the two preceding figures. The GMR rises significantly up to
$50$ \% impurities - i.e.\ this time we have the same trend for both models
(i) and (ii) --, and then it falls again, which is natural since a system of
two {\it three-ML} ferromagnets separated by a 4-ML-spacer and $x>50$ \%
interface impurities is equivalent with two {\it four-ML} ferromagnets
separated by a 2-ML spacer with $x'\,=\,(1-x) < 50$ \% interface impurities.
Here it should be noted that the majority-spin carriers are hardly
influenced by our interface disorder, which is clear since we have assumed
that the s-bands and the majority d-bands of our metallic ferromagnets and
of the metallic spacers are identical. But it should also be noted that
 at the above-mentioned maximum  the
CIP-GMR with {\it interface} scattering reaches much higher values (e.g.\ 9\%
and 8\% respectively, for the parameter sets of (i) and (ii)) as with
our nonmagnetic impurities in the {\it bulk}.
\section{Non-metallic spacers, CPP geometry}
Finally, we come to our most startling result, for the non-metallic spacers,
where we have considered metallic ferromagnets as in model (i), but
{\it {{semi-conducting}} spacer layers} as in the first line of Tables 3 and 4. The
geometrical situation corresponds to the CPP geometry of Fig.\ 1, with
metallic ferromagnets (two-times 3 monolayers) and a variable number $n_s$
of semi-conducting
 spacer layers, i.e.\ with a conduction band and a valence band, inbetween. Of course one now expects a strong exponential
decrease of all conductances with increasing $n_s$, but GMR
=($\Gamma^{++}/\Gamma^{+-})\, -\,1$ may be well defined and even {\it
increase} with increasing $n_s$, as long as the  spacer thickness does not
become larger then the 'dephasing scattering length' or 'spin-flip
scattering length'.

The calculations have been performed by the highly accurate $\vec k$-space
method in reliable numerical accuracy, and we have plotted the conductances
and the GMR against the number
$n_s$ of spacer layers, calculated in the ballistic limit of perfect phase
coherence. In Fig.\ 18 and Fig.~19 there is now the expected drastic exponential
decrease of the conductance, 	
which for our model
is {\it weaker} for $\Gamma^{++}$ than for $\Gamma^{+-}$. (The decay is
weakest for $\Gamma^{++}_\downarrow$, which is reasonable, since the
 $\downarrow$ d-electrons have a much lower energy gap to the lower edge of
the semi-conductor conduction band than the $\uparrow$ d-electrons,
 and with the $\Gamma^{+-}$ case, in contrast to the $\Gamma^{++}$ case, there
 are the additional 'matching' problems at the interface to the second 
ferromagnet. Note
that due to the s-d-hybridization, and with the finite value of $t_d$, the 
d-electrons do in any case contribute to the tunneling current.)
 So this means that in our
case, and in the recent calculation of MacLaren {\it et al}, \cite{Butler2},
 to be discussed below, and
 in contrast to Julliere's assumptions, \cite{Jul},  the
tunneling matrix through the semi-conducting spacer is naturally strongly
spin-dependent. (The fact that here for $\Gamma^{+-}$ the exponential decay
with the spacer thickness is stronger than for $\Gamma^{++}$, depends of
course on our model:
for other assumptions on the semi-conductor it
might be just the other-way round.)

  As a consequence, one gets a
positive GMR, which {\it strongly increases with} $n_s$, e.g.~up to GMR $\sim 300$
\% for $n_s\widetilde > \,5$ in case of model (iv), Fig.\ 18, i.e.~for the
model with the 'direct energy gap'. For the case of model (iii) with the
'indirect gap', Fig.\ 19, the increase is at first similar, i.e.~for $n_s
\widetilde < 4$, but then the GMR increases even further to {\it collosal}
values, e.g.~GMR $\sim 2500$ \% for $n_s=7$. However because of the strong
exponential decrease of the conductance, and in view of the fact that we
have neglected indirect transitions involving phonons, this region is
probably beyond experimental realizability, and the results of Fig.\ 18 and
Fig.\ 19 should not be taken literally, although the numerical accuracy of
our calculations seems to be sufficient  (In Fig.\ 18 and Fig.\ 19, the Cunningham
accuracy parameter, see \cite{Cun}, was $m=10$).

In fact, however, in the already cited recent paper of MacLaren {\it et al.},
\cite{Butler2}, which contains an ambitious realistic LKKR calculation of
the ballistic CPP conductances and the corresponding GMR for bcc (001)
Fe/ZnSe/Fe trilayers, the conductances of Fig.\ 3 of that paper resemble
very much -- almost quantitatively -- to the results of our simple model
calculation in Fig.19. One only must take into account that in the paper of
MacLaren {\it et al.}, one Zn-Se bcc double-layer is replaced by two monolayers in
our simple {\it sc} two-band model calculation with the parameter set (iii).
So here we have an example where a fully realistic ab-initio calculation may be
sometimes not necessary after all.

 Finally we present at this place additional comments  on the seminal
paper of Julliere, \cite{Jul}, written already in 1975. Julliere describes the
tunneling conductance from the  ferromagnetic metallic {\it  source} to a metallic
ferromagnetic {\it drain} through an isolating Al$_2$O$_3$ spacer as a
product of  the densities of states (DOS) of the ferromagnetic 
source and drain, multiplied by an
effective tunneling matrix element. This should be essentially equivalent to
our approach, which involves two imaginary parts of Green's function, which
 represent in principle densities of states. But from the relevant densities
of states of the magnetic metals, the contributions of {\it confined} electrons, which
contribute to the DOS, but not to the conductance, should be excluded; such
confined states exist e.g.\
 with the
Co down-spin electrons, which are strongly reflected at the Cu interfaces;
so instead of Juliere's DOS, an 'effective DOS' should be used, and since the
reduction  factor of the DOS to this 'effective DOS' is spin-dependent, the
remaining 'effective tunneling matrix elements', if one uses the original
DOS, i.e.\ Julliere's formula,
would in any case depend on the spin, contrary to Julliere's assumption,
 if applied to systems as Co/Al$_2$O$_3$/Co, \cite{MathonPriv}. 

In principle, however, our approach is even more explicit, and more demanding,
 since it also involves the
transfers to (and  from) the ideal leads to the ferromagnetic metal, and
from the reservoirs to the ideal leads.

In Julliere's  TMR experiments, and similar experiments performed at
present, the high values of the TMR
predicted in the present paper and also in the paper of MacLaren {\it et al.},
\cite{Butler2}, have {\it not} been observed hitherto. However one should stress
that our insulating layers are extremely thin (from 1 to 15 monolayers only)
with ideal interfaces, whereas the experimental thicknesses of the
Al$_2$O$_3$ spacers are larger and with rougher interfaces, so that in that case
spin-channel mixing might play a role, particularly if on the metallic side
of the interface the transverse spin components are almost as probable as the
$z$-components. The band-structure, e.g.\ that of the semi-conductors
 involved, is also crucial, since the
exponential decay with the spacer-thickness should be weaker for $\Gamma^{++}$
than for $\Gamma^{+-}$, i.e.\ the semi-conducting compounds  involved should have
 properties similar to those of our present models, or to the Fe/ZnSe/Fe-system
 studied by Mac\-Laren {\it et al.}, \cite{Butler2}. In any case, one can only
 speculate that perhaps ballistic point contacts as 'nanocontacts' might
ultimately work, with particular
 few-$\AA$-thick
semi-conducting nanospacers, and with
very clean interfaces.

The suggested experiment concerning the TMR
 would therefore be analogous to the very recent set-up of Garcia {\it et al.},
\cite{Garcia}, but with a small semi-conducting tip of the above-mentioned
kind.

  \section{Conclusions}
We have performed a systematic study of  the CPP-GMR and CIP-GMR (i.e.\ with
{\it Current Perpendicular to the Planes} or {\it Current in the Planes},
respectively), for systems consisting of two 'ideal leads' attached to two
ferromagnetic metallic slabs, which are three monolayers thick, separated by
$n_s$ monolayers of non-magnetic (metallic or non-metallic) spacer material.
A simple two-band model is used throughout, where in the metallic case
 one band mimics the s-bands, and the other one   the
d-bands
 which are spin-dependent in the 
ferromagnetic metals, whereas for nonmetallic spacers the
two bands describe
 the valence band and the conduction band, respectively. The
s-d-hybridization is taken into account, and is important for our results.
For the metals, and also for the semi-conducting spacer, we use two
{{specific}} parameter sets,
 corresponding in the metallic case very roughly to the situation
of a Co/Cu/Co-trilayer system with Cu leads, whereas in the non-metallic
case the two parameter sets of the spacer distinguish between situations
corresponding to an 'indirect' and a 'direct' energy gap, respectively.

The calculations have been performed by very accurate
 Green's function methods and
applied to systems without impurities, with {{{nonmagnetic}} 
bulk impurities in the spacer,
and with {{magnetic and non-}} {{magnetic}} interface impurities
 produced by mutual interdiffusion of the atoms near the
interface. Within our formalism other situations could also have been
treated, e.g.\ more complicated geometries as (i) the 'constrictions'
studied  experimentally by Garcia {\it el al.}, \cite{Garcia}, or (ii)
 quasi-one-dimensional problems, where the potentials are
constant within a layer, but random from one layer to the next. The
last-mentioned problem has already been extensively studied by J.\ Mathon,
\cite{Mathon97}, and in two-dimensional systems by one of the present
authors, \cite{Boehm99}. Also the recent work of Sanvito {\it et al.} should
be mentioned in this broader context. 

 In all cases the influence of the strength of
disorder leads to strong effects, sometimes to an enhanced GMR and sometimes
to a reduction. However for different model parameters the outcome can be
quite different, which is not astonishing after all, since three
conductances are involved in the changes, namely $\Gamma^{(++)}_\uparrow$,
 $\Gamma^{(++)}_\downarrow$, and $\Gamma^{(+-)}_\sigma$, and the GMR effect is --
after all -- a difference effect out of these quantities. 

So our
 calculation shows among other results that for the
GMR one can hardly avoid extensive calculations for very realistic models of
the specific systems considered.
Only for the CPP case with impurities in the bulk of a metallic spacer, the GMR
seems to be strongly reduced with increasing strength $\sigma_r$ of the
disorder, but only as long as $\sigma_r$ remains $\widetilde < \, 1$ (Fig.\ 8
and Fig.\ 9). In contrast, interface impurities may lead in the same case to
a reduction of the GMR for one parameter set (Fig.\ 10), but to an
enhancement for another set (Fig.\ 11).

 One of our main results is perhaps the {{suggested}} existence of
a drastic {\it increase} of the GMR with increasing spacer thickness in case
of ballistic conductance through a {\it nonmetallic} spacer, see section 4
above, 
where we also  suggest an experimental realization
along the lines of the recent experiments of Garcia {\it et al.},
\cite{Garcia}.
 
\subsection*{Acknowledgements}
We would like to thank Prof.\ Dr. P.\ Bruno, 
Prof. Dr. P. Weinberger, and Prof.\ Dr.\ G.\ Bayreuther, {{and the
experimentalists around him}}, for stimulating discussions.
S.K. thanks the Humboldt foundation for financial support and the University
of Regensburg for its hospitality, and U.K.  does the same  w.r.\ to  the
Polish Academy of Sciences and the Institute of Molecular Physics  in
Pozna\'n.

\newpage

\centerline{\bf{Tables}}

\begin{table}[ht]
\begin{center}
\begin{tabular}{|l|c|c|c|c|}\hline
\multicolumn{5}{|c|}{Model parameters (i), from overall
band-structure}\\\hline
&\multicolumn{2}{|c|}{\rule{1.8cm}{0cm}Co\rule{1.8cm}{0cm}}&\multicolumn{=
2}{c|}{Cu}\\\cline{2-5}
\raisebox{1.5ex}[-1.5ex]{}& E(s) & E(d,$\sigma$) & E(s) & E(d) \\\hline
$\uparrow$  &0.0&-2.6&0.0&-2.6\\\hline
$\downarrow$&0.0&-2.1&0.0&-2.6\\\hline
\multicolumn{5}{|l|}{Fermi energy: $E_F=3D-2.8$}\\
\multicolumn{5}{|l|}{Hopping integrals: $t_s=3D-1, t_d=3D-0.2,
t_{sd}=3D0.3$}\\\hline
\end{tabular}
\end{center}
{\sf\caption[Model parameters (i)]{Model parameters (i); metallic
spacer}
\label{t:modelparameter1}}

\vskip3mm
\begin{center}
\begin{tabular}{|l|c|c|c|c|}\hline
\multicolumn{5}{|c|}{Model parameters  (ii), from overall behaviour of
the
 DOS}\\\hline
&\multicolumn{2}{|c|}{\rule{1.8cm}{0cm}Co\rule{1.8cm}{0cm}}&\multicolumn{=
2}{c|}{Cu}\\\cline{2-5}
\raisebox{1.5ex}[-1.5ex]{}& E(s) & E(d,$\sigma$) & E(s) & E(d) \\\hline
$\uparrow$  &0.0&-1.0&0.0&-1.0\\\hline
$\downarrow$&0.0&-0.2&0.0&-1.0\\\hline
\multicolumn{5}{|l|}{Fermi energy: $E_F=3D0.0$}\\
\multicolumn{5}{|l|}{Hopping integrals: $t_s=3D-1, t_d=3D-0.2,
t_{sd}=3D1.0$}\\\hline
\end{tabular}
\end{center}
{\sf\caption[Model parameters (ii)]{Model parameters (ii);
  metallic spacer}\label{t:modelparameter2}}
\end{table}

%
\begin{table}[ht]
\begin{center}
\begin{tabular}{|l|c|}\hline
&'Isolator 1'\\\cline{2-2}
&$t_s\equiv -1.0, t_d\equiv -0.2, t_{sd}\equiv 0.3$\\\hline
small gap~: \hfill\quad $0.025eV$ & $E_s=-8.77$\hfill$E_d=-1.60$ \\\hline
intermediate gap~:\hfill\quad $ 0.17eV$ & $E_s=-8.85$\hfill$E_d=-1.55$ \\\hline
large gap~:\hfill\quad $  3.4eV$ & $E_s=-9.77$\hfill$E_d=-0.60$ \\\hline
\end{tabular}
\end{center}
{\sf\caption[Parameter set (iii), first line, with 'indirect' Energy Gap]
{Parameter set (iii): Non-metallic spacer;   'indirect energy gap';
$E_F=-2.8$; the set of the first line was always used in the paper,
with $t_s\equiv -1$, $t_d\equiv -0.2$, $t_{sd}\equiv 0.3$ throughout.
}\label{t:isolator1}}
\end{table}

\newpage


\begin{table}[ht]
\begin{center}
\begin{tabular}{|l|c|}\hline
&'Isolator 2'\\\cline{2-2}
&$t_s\equiv -1,\, t_d=0^{*)}, t_{sd}\equiv 0.3$\\\hline
energy gap at $k_{[111]}=0.814$ :\hfill\quad $1.0eV$ & $E_s=2.50$\hfill$E_d=-2.85$ \\\hline
energy gap at $k_{[111]}=1.888$ :\hfill\quad $1.0eV$ & $E_s=0.00$\hfill$E_d=-2.80$ \\\hline
energy gap at $k_{[111]}=3.066$ :\hfill\quad $1.0eV$ & $E_s=-4.0$\hfill$E_d=-2.80$ \\\hline
\end{tabular}
\end{center}
{\sf\caption[Parameter set  (iv), first line, corresponding to a 'direct
gap']
{Parameter set (iv):  non-metallic spacer; 'direct energy gap';
$E_F=-2.8$; the arrow in Fig.\ 5 denotes the position of the minimal energy difference
$E_c(\vec k)-E_v(\vec k)$; the set of the first line was always used
in the paper; $t_s\equiv -1$ and $t_{sd}\equiv 0.3$  throughout, as
before,$^{*)}$ and
$t_d=-0.2$  for the hopping between two neighbouring metal atoms; whereas, if at least
one of the two atoms were  nonmetallic, this time $t_d=0$ was assumed, so that the
dispersion of the valence band in the r.h.s.\ of Figure\ 5 is solely
caused by $t_s\equiv -1$ and the on-site s-d-hybridization $t_{sd}\equiv 0.3$.
 \label{t:isolator2}}}
\end{table}
\vfill \eject

\centerline{\bf{Figures}}

\epsfxsize=14cm
\epsfbox{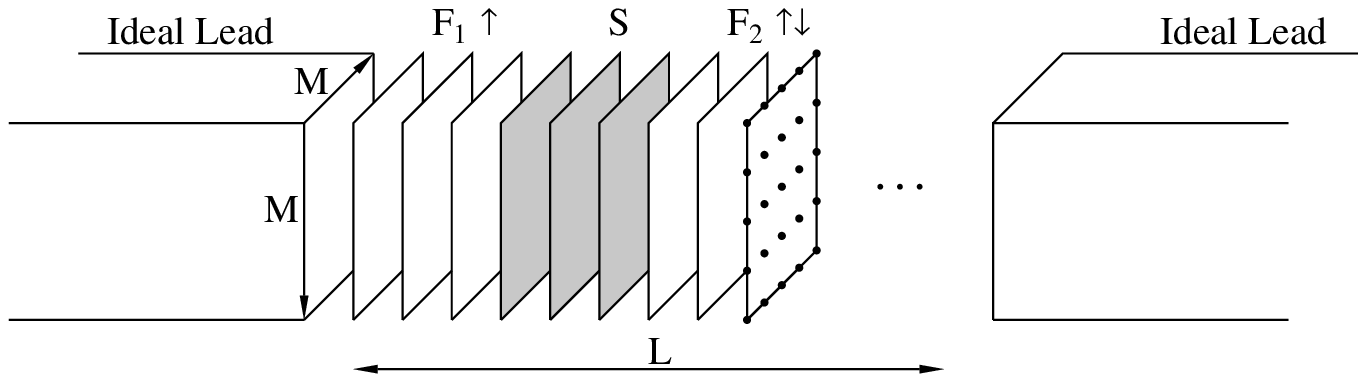}

\noindent Figure 1:
{{{Schematic plot of our systems for the CPP-geometry}}}

\vglue 1 truecm
\epsfxsize=14cm
\epsfbox{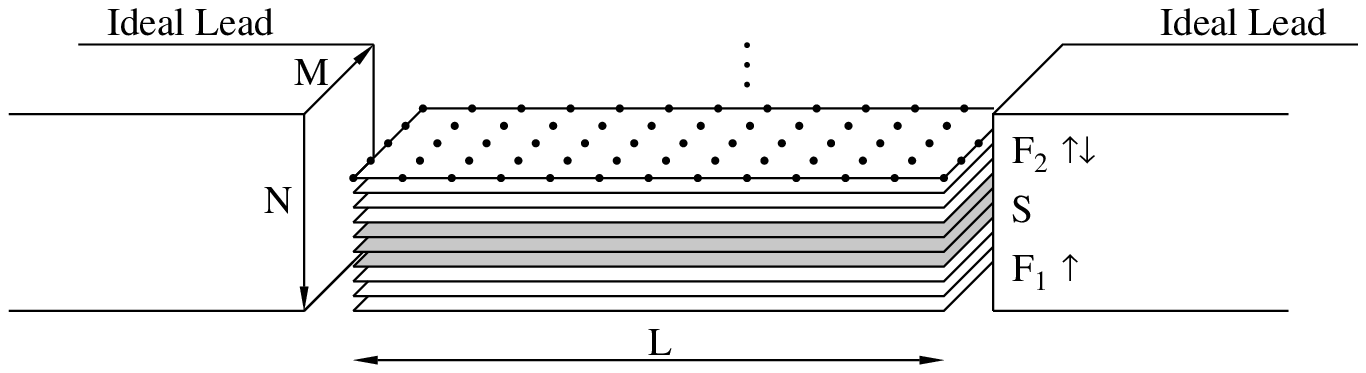}

\noindent Figure 2:
{{{Schematic plot of our systems for the CIP-geometry}}}

\epsfxsize=12cm
\epsfbox{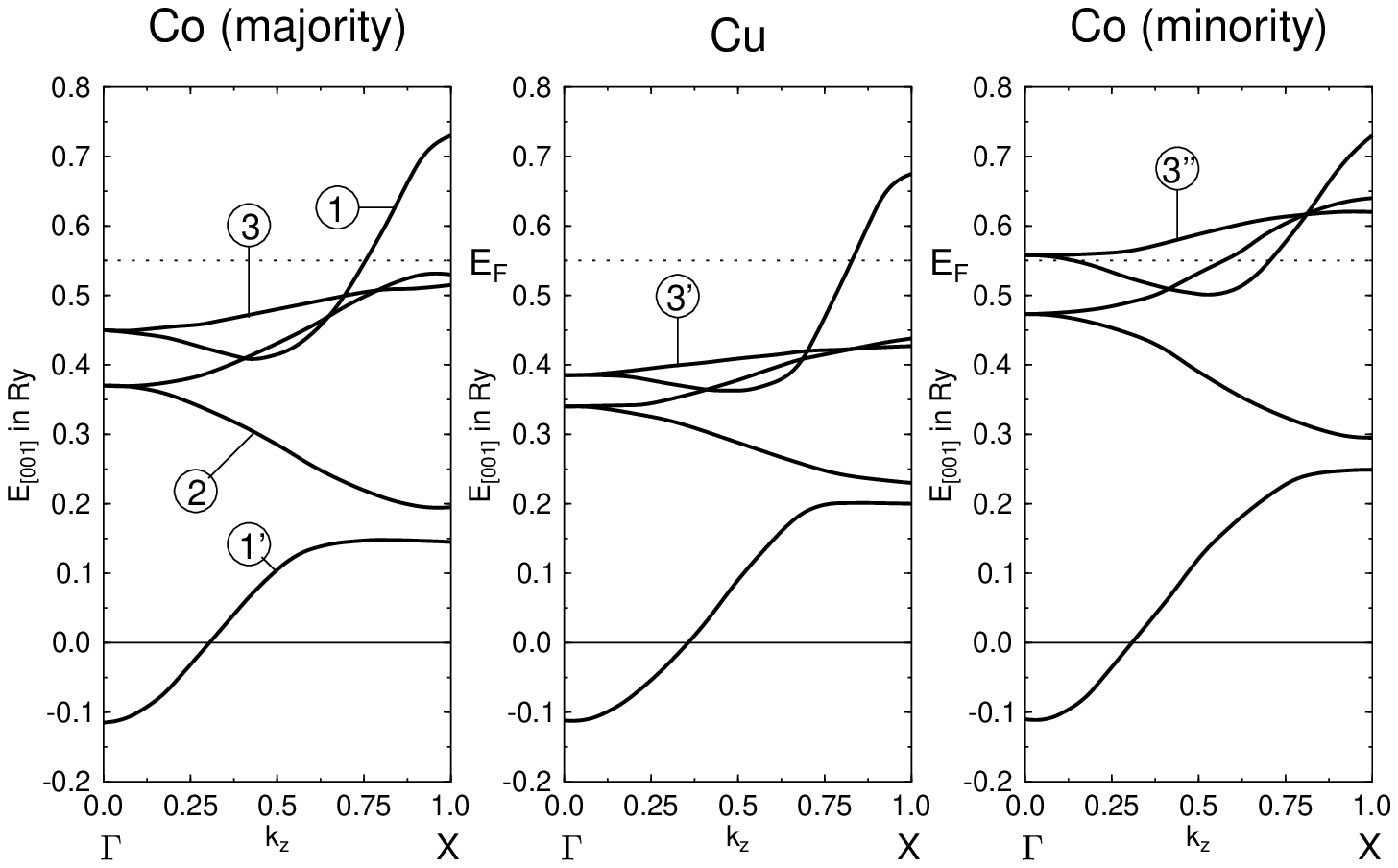}

\noindent Figure 3:
{ Realistic band-structure of Co and Cu
 according to \cite{Mat3}.
 Note the similarity of Cu and Co$_\uparrow$ bands}

\vglue 1 truecm
\epsfxsize=14cm
\epsfbox{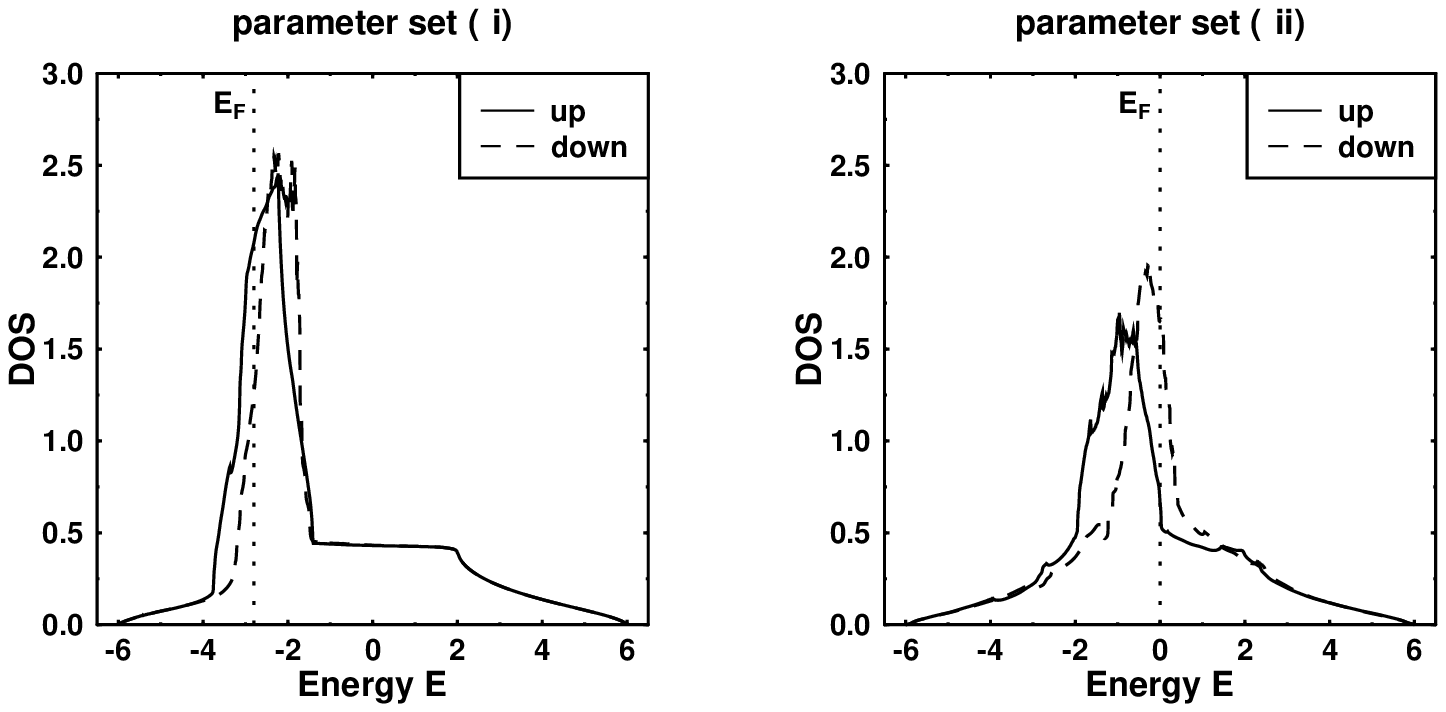}

\vglue 0.5 truecm
\noindent {\centerline{Figure 4: 
{Densities of
states for the magnetic metals, for models (i) and (ii)}}}

\epsfxsize=14cm
\epsfbox{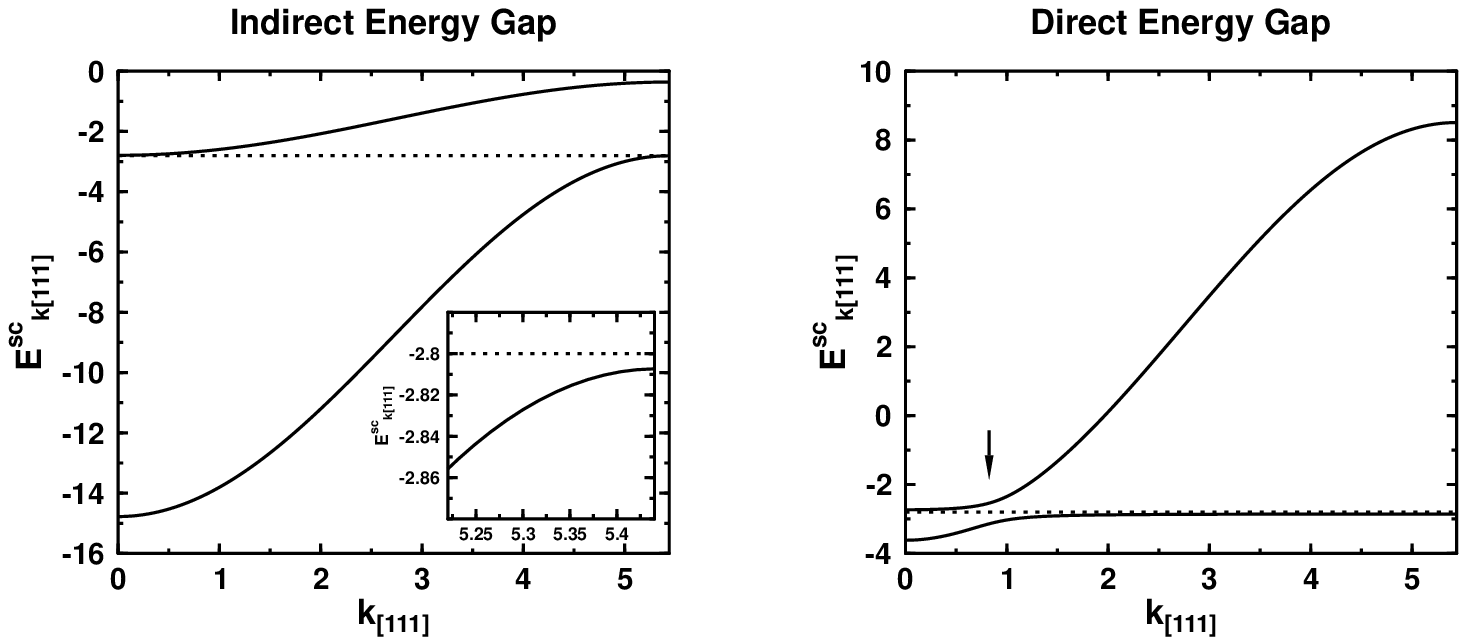}

\vglue 0.5 truecm
\noindent Figure 5:
 {Band-structure of the non-metallic
spacers (iii) and (iv); $E_F=-2.8$; on the left h.s.~the band structure of
model (iii) is shown, with an 'indirect energy gap' from   the upper edge
of the valence band at $k_{\rm 111}=\sqrt{3}\pi$ to the lower edge of the
conduction band at $\vec k =0$, which is two times the energy distance to
$E_F$  (see the inset); the r.h.s.~is for model (iv)
with a 'direct gap' represented by the arrow.}

\vglue 1.5 truecm
\epsfxsize=14cm
\epsfbox{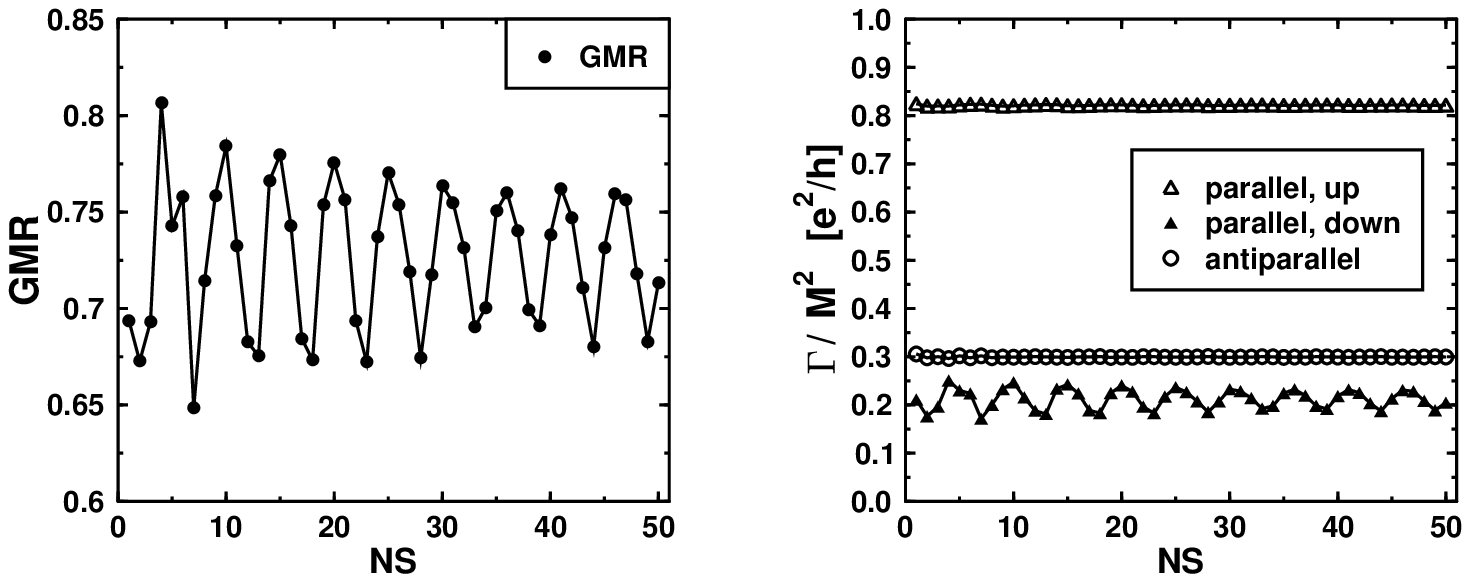}

\vglue 0.5 truecm

\noindent{\centerline{ Figure 6:
{k-space calculation, CPP, system (i)}}}

\epsfxsize =14cm
\epsfbox{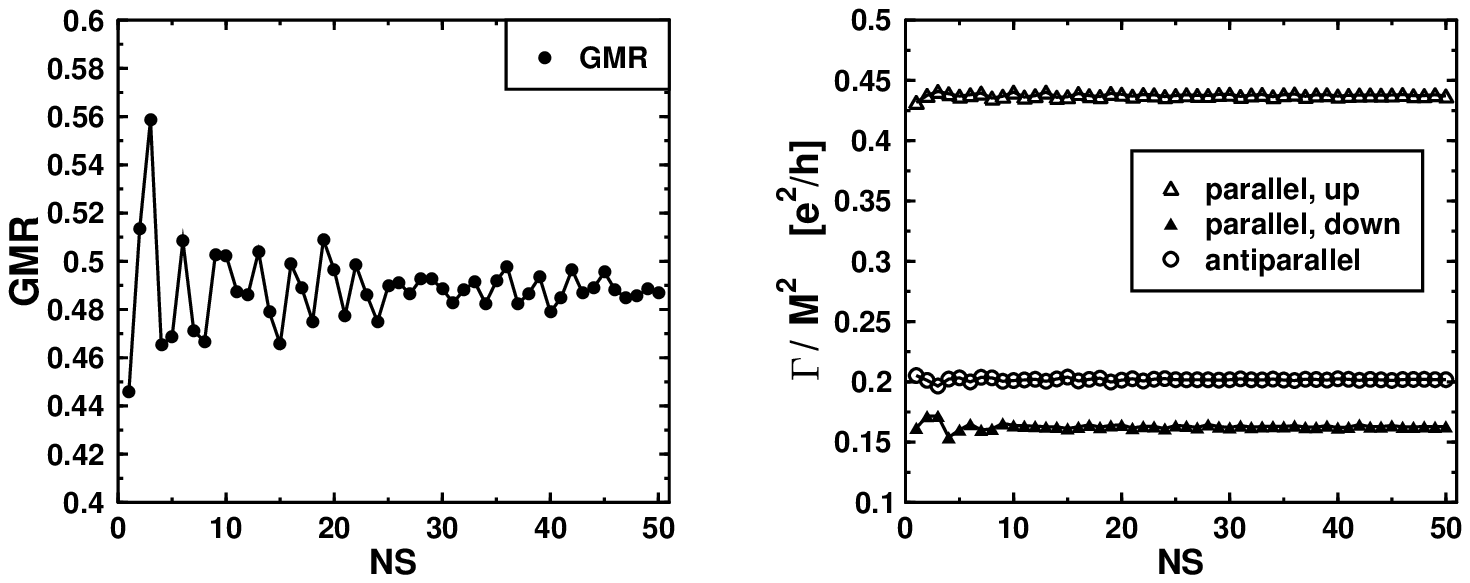}

\vglue 0.5 truecm
\noindent {\centerline{Figure 7: 
{k-space calculation, CPP, system (ii)}}}

\vglue 2.0 truecm

\epsfxsize=14cm
\epsfbox{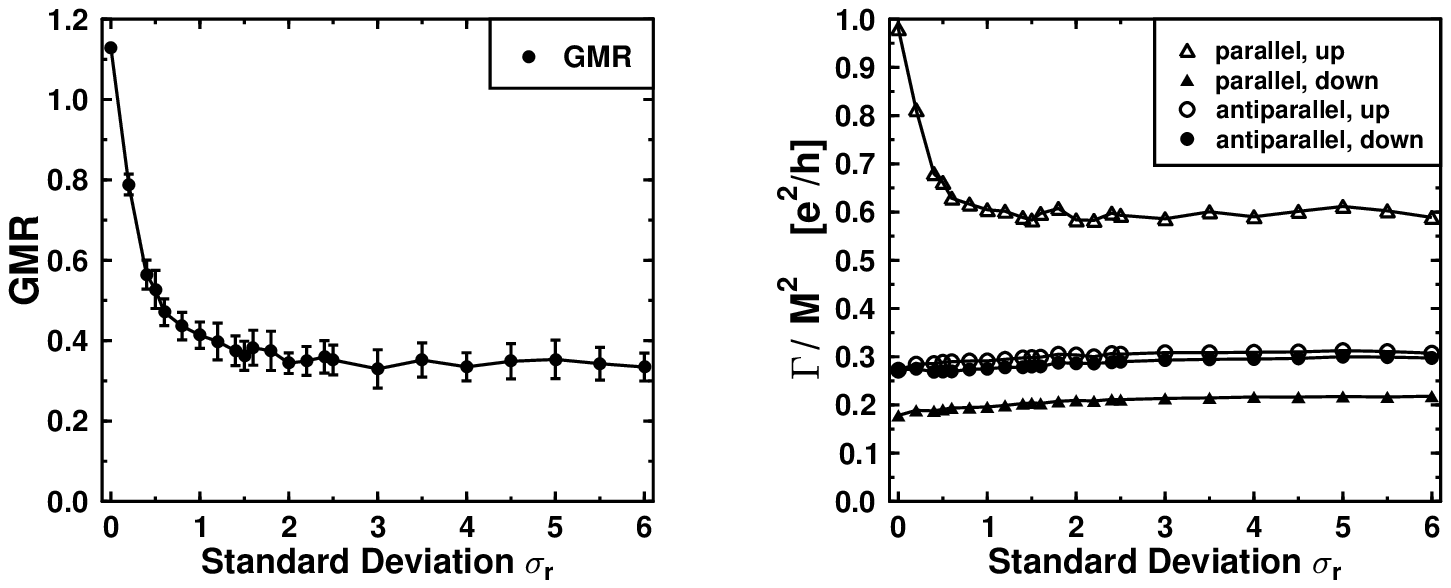}
\vglue 1.0 truecm
\noindent Figure 8:
{x-space calculation, CPP, system (i), bulk
impurities; the standard-deviation $\sigma_r$ of Eq. (\ref{eqBulk_impurities})
characterizes the 'strength of disorder'; 20 samples per point}

\epsfxsize=14cm
\epsfbox{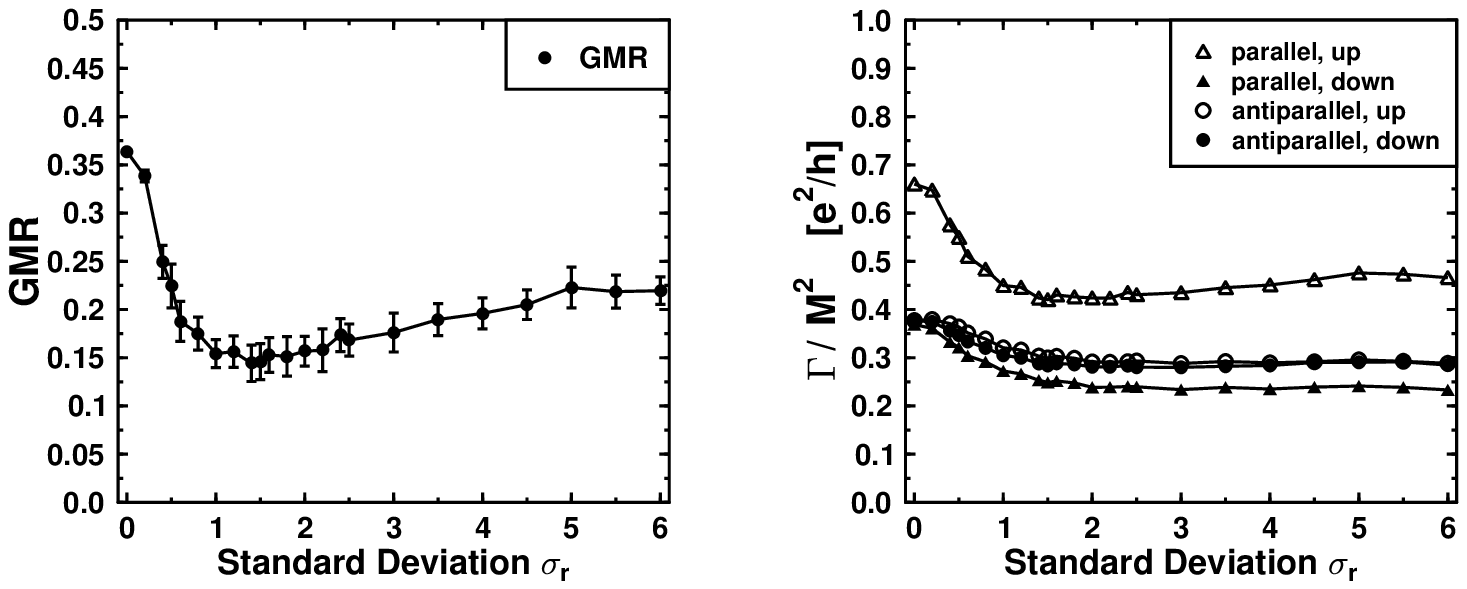}
\vglue 1 truecm

\noindent Figure 9: 
{x-space calculation, CPP, system (ii), bulk
impurities;  the standard deviation $\sigma_r$ of Eq. (\ref{eqBulk_impurities})
characterizes the 'strength of disorder'; 20 samples per point}
\vglue 1.5 truecm

\epsfxsize=14 cm
\epsfbox{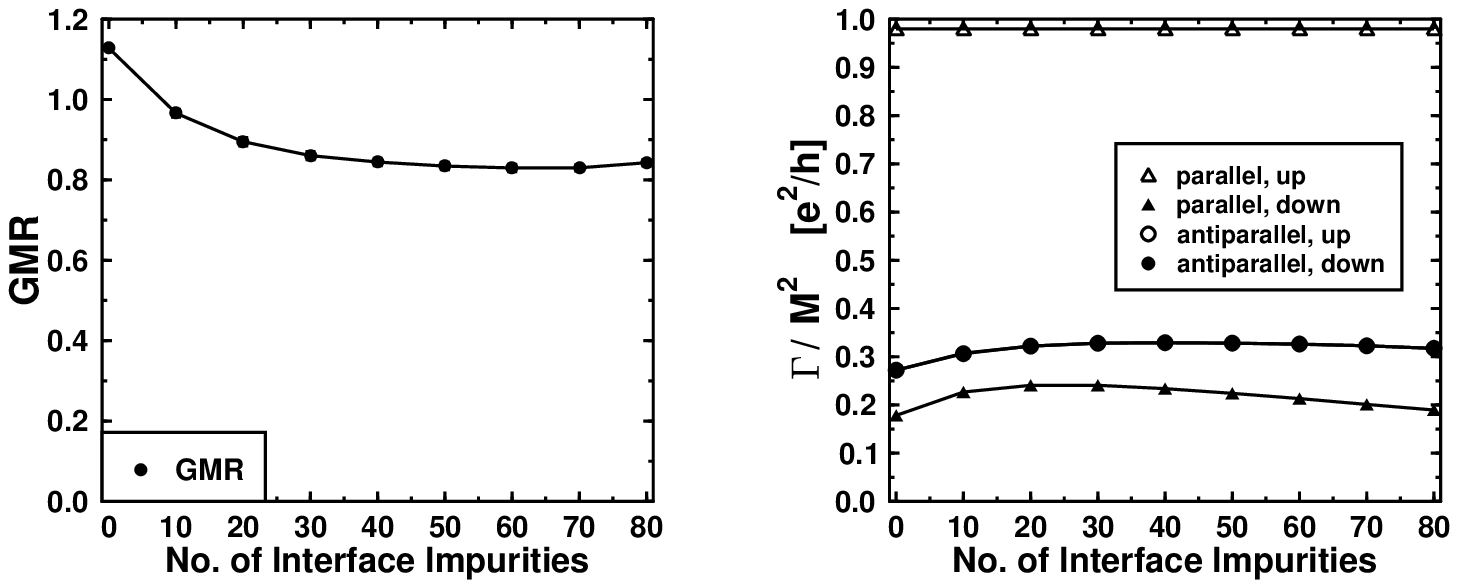}
\vglue 1 truecm
\noindent Figure 10: 
{x-space calculation, CPP; system (i); interface
impurities; 100 samples per point}

\epsfxsize=14 cm
\epsfbox{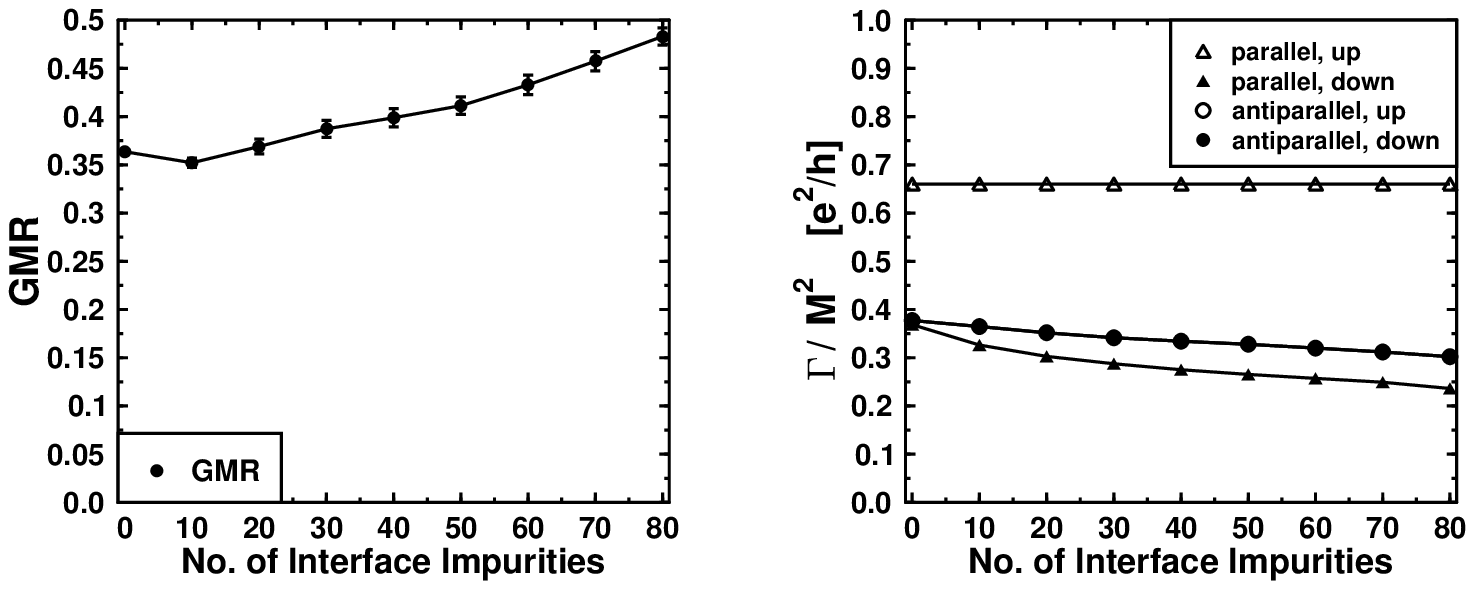}
\vglue 1 truecm

\noindent Figure 11:
{x-space calculation; CPP; system (ii);
 interface impurities; 100 samples per point}

\vglue 1.5 truecm

\epsfxsize=14cm
\epsfbox{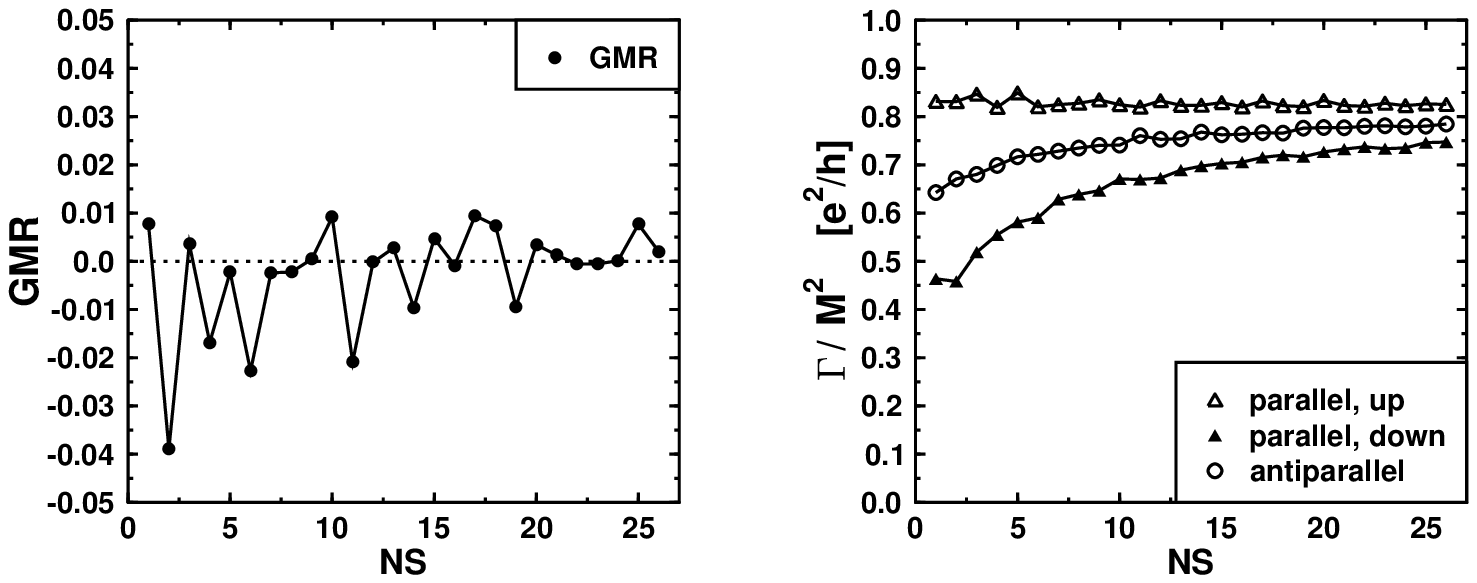}
\vglue 1 truecm
\noindent {\centerline{Figure 12: 
{x-space calculation, CIP, system (i); no impurities}}}

\vglue 1.5 truecm
\epsfxsize=14cm
\epsfbox{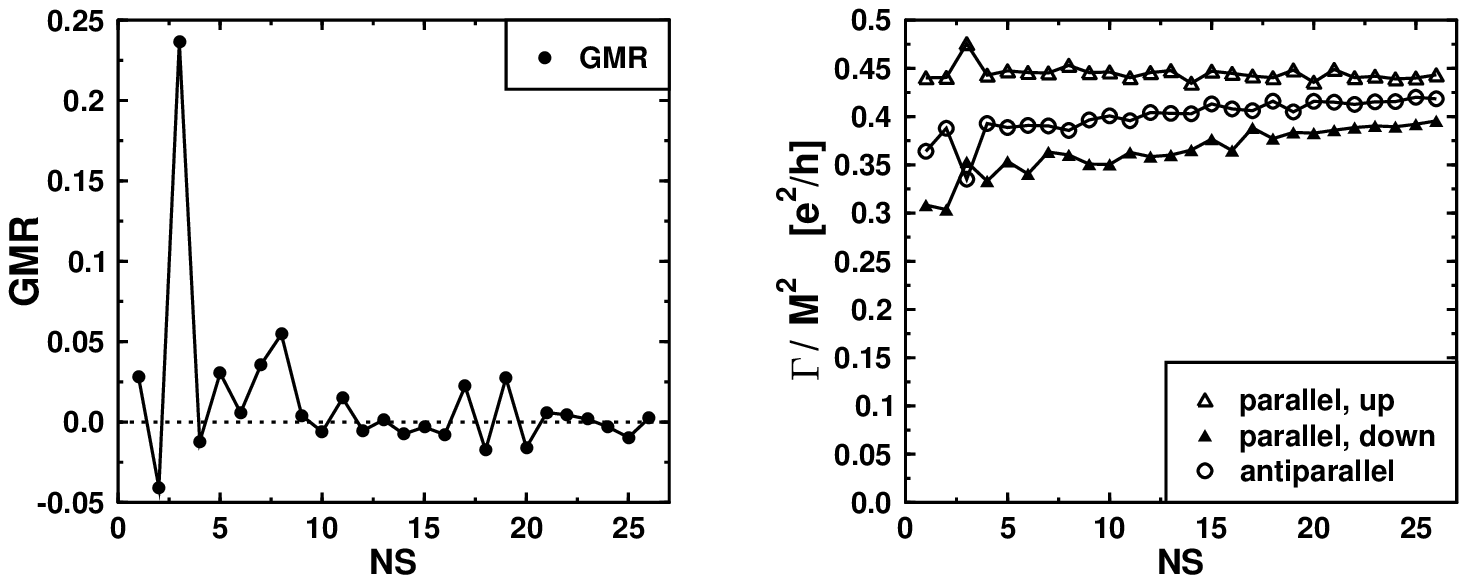}
\vglue 1.0 truecm

\noindent{\centerline{Figure 13:
{x-space calculation, CIP, system (ii); no impurities}}}

\epsfxsize=14 cm
\epsfbox{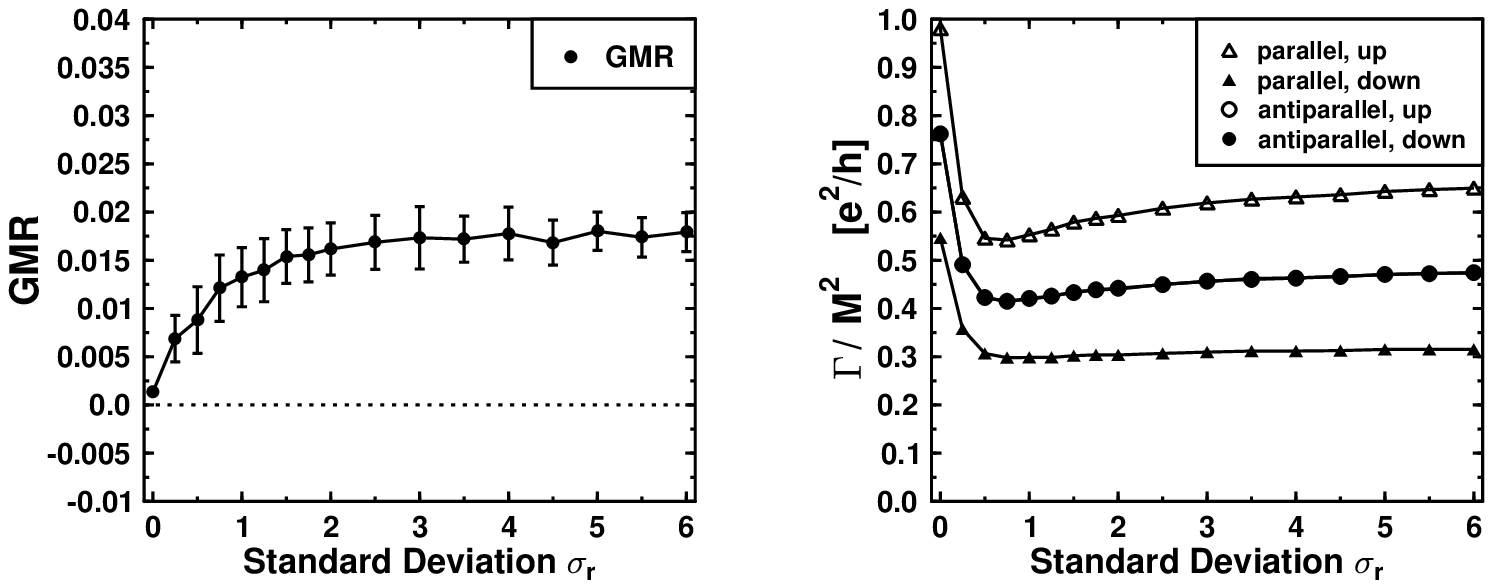}

\vglue 1 truecm

\noindent Figure 14: 
{x-space calculation; CIP; system (i); bulk
impurities;  the 'standard deviation' $\sigma_r$ of Eq. (\ref{eqBulk_impurities})
characterizes the 'strength of disorder'; 100 samples per point}

\vglue 2.5 truecm
\epsfxsize=14 cm
\epsfbox{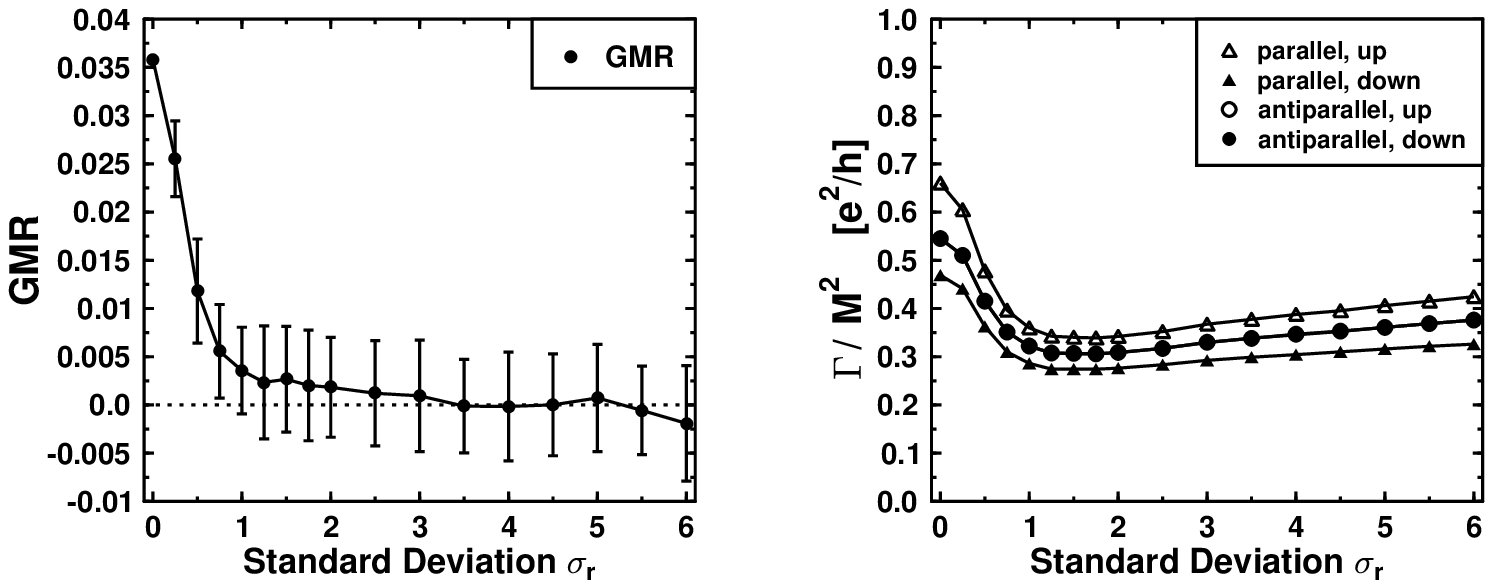}

\vglue 1 truecm
\noindent Figure 15: 
{x-space calculation; CIP; system (ii); bulk
impurities;  the 'standard deviation' $\sigma_r$ of Eq. (\ref{eqBulk_impurities})
characterizes the 'strength of disorder'; 100 samples per point}

\epsfxsize=14 cm
\epsfbox{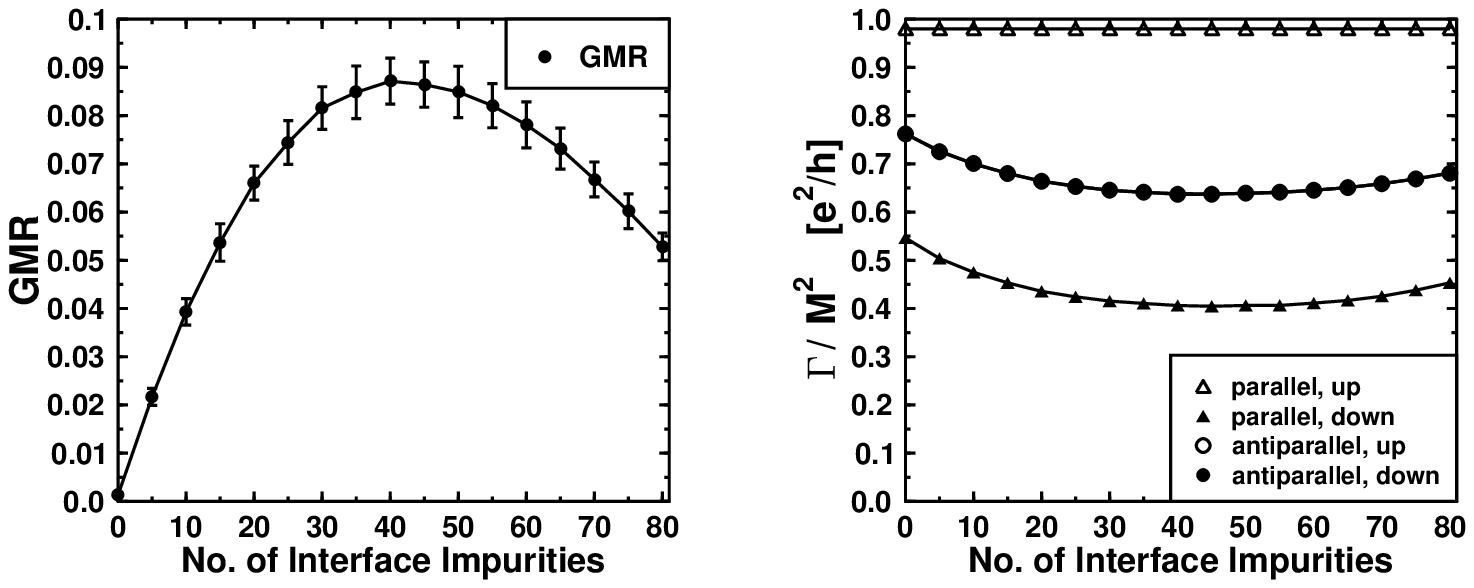}

\vglue 1 truecm

\noindent Figure 16: 
{x-space calculation; CIP; system (i); interface
impurities; 100 samples per point}

\vglue 1.5 truecm
\epsfxsize=14 cm
\epsfbox{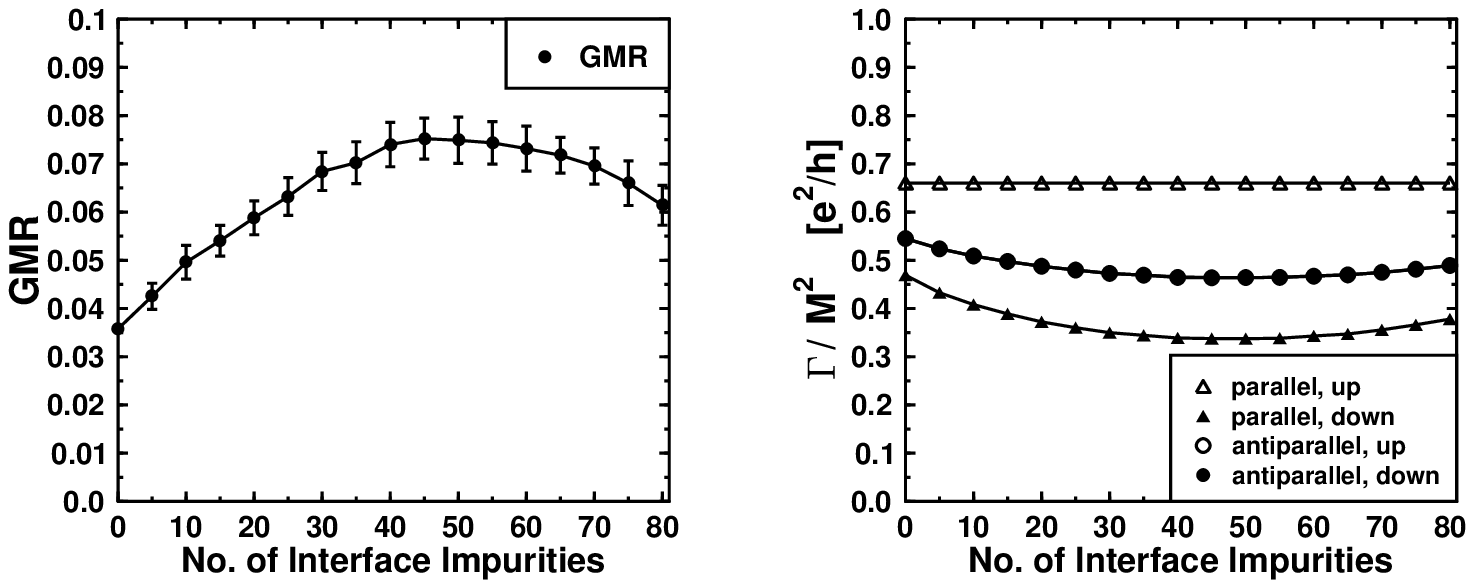}
\vglue 1 truecm

\noindent Figure 17: 
{x-space calculation; CIP; system (ii); interface
impurities; 100 samples per point}

\epsfxsize=14cm
\epsfbox{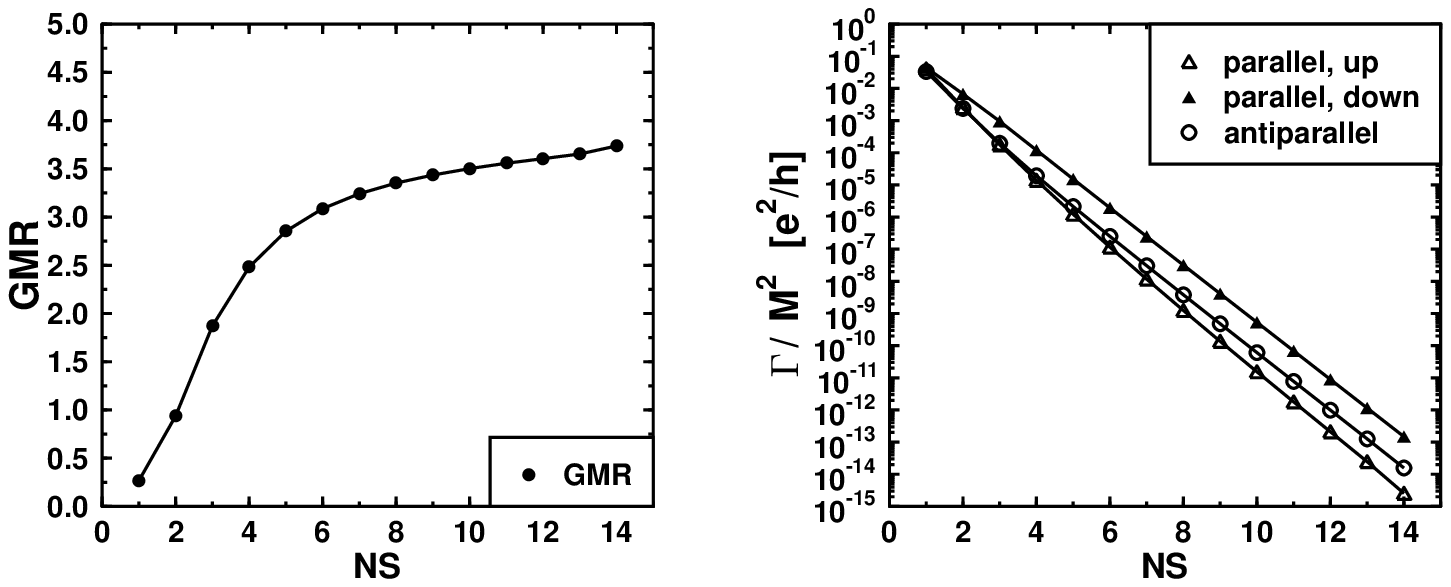}
\vglue 1.0 truecm
\noindent Figure 18: 
{CPP-GMR for non-metallic spacer (iv), i.e.~with {\it direct} gap; 
 $NF_1=NF_2=3$; parameters corresponding to line 1 in Table 4; no
impurities}

\vglue 1.5 truecm
\epsfxsize=14 cm
\epsfbox{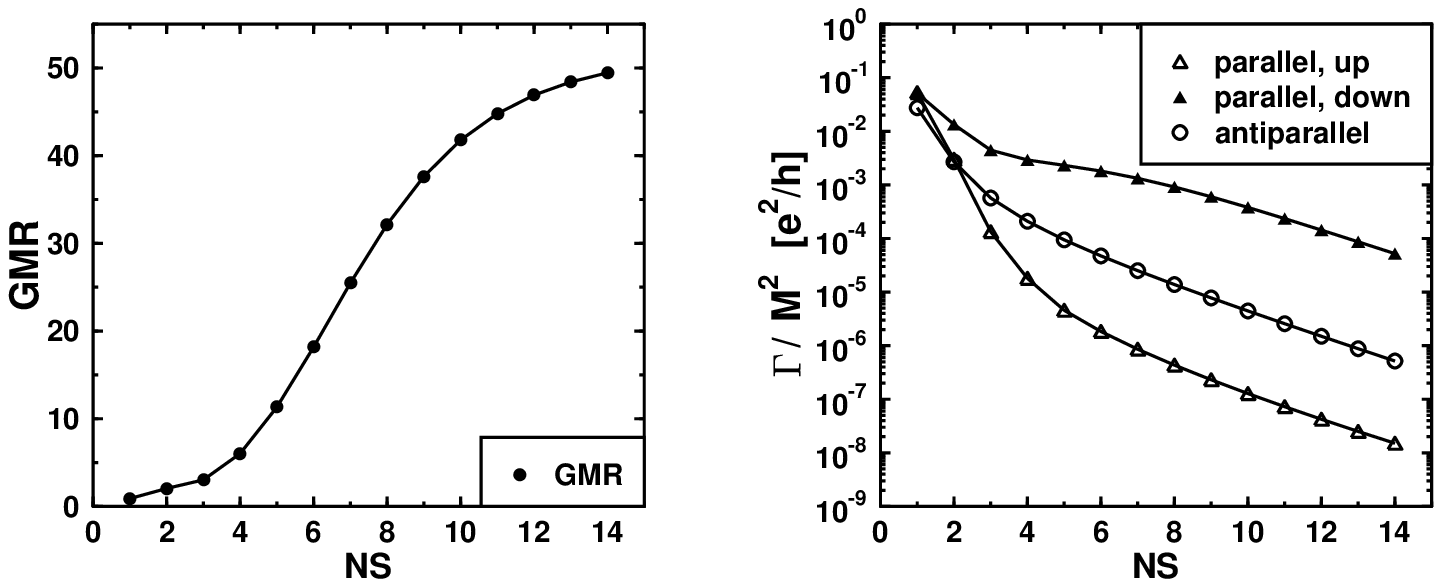}
\vglue 1 truecm

\noindent Figure 19:
{CPP-GMR for non-metallic spacer (iii), i.e. with {\it indirect} gap; 
$NF_1=NF_2=3$; parameters corresponding to line 1 in Table 3; no
impurities}


\begin{thebibliography}{99}
\bibitem{REMrealspace}
 Here the following
remarks are in order:\newline
 a) Calculations have also been performed with 8$\times$8 and
16$\times$16 lattices, but our 10$\times$10-systems with free boundary
conditions in the transverse directions, and ideal leads -- i.e.\ infinite
extension in the longitudinal direction -- turned out to be the
optimal compromise between excessive computing time and quality of the
results.\newline b) Although we consider only single-site impurity potentials, every
scattering wave can propagate in all directions, i.e.\ no component of $\vec
k$ is generically conserved in the scattering.
\end{thebibliography}
\end{document}